
\magnification=\magstep1
\baselineskip 8 true mm
\def\ni{\noindent}
\def\laq{\raise 0.4ex\hbox{$<$}\kern -0.8em\lower 0.62 ex\hbox{$\sim$}}
\def\gaq{\raise 0.4ex\hbox{$>$}\kern -0.7em\lower 0.62 ex\hbox{$\sim$}}

\rightline{CERN-TH.7130/93}
\vskip 1.5cm
\centerline{\bf General Cosmological Features of the
Einstein--Yang--Mills--Dilaton System}

\centerline{\bf in String Theories }

\vskip 1.0cm

\centerline{M. C. Bento\footnote{$^*$}{On leave of absence from Departamento de
F\' \i sica, Instituto
 Superior T\'ecnico, Av. Rovisco Pais, 1096 Lisboa Codex, Portugal}
 and O. Bertolami$^*$}

\vskip 0.5cm
\centerline{\it CERN, Theory Division}
\centerline{\it CH-1211  Geneva 23}
\centerline{\it Switzerland}

\vskip 1.0cm

\centerline{ABSTRACT}

\vskip 2mm

We establish the main features of homogeneous and isotropic dilaton,
metric and Yang-Mills field configurations in a cosmological
framework. Special attention is paid to the energy exchange between
the dilaton and the Yang-Mills field and, in particular, a new
 energy exchange term is identified. Implications for the
Polonyi problem in 4-dimensional string models and in dynamical
supersymmetry breaking scenarios are discussed.

\vskip 1.5cm
CERN-TH.7130/93

February 1994

\vfill
\eject

Although string theory is the best candidate advanced so far to make gravity
compatible with quantum mechanics and  unify all the fundamental
interactions of nature, this unification takes place
at very high energy, presumably at the Planck scale, and it is very
difficult to extract unambiguous predictions for physics at the energy
scales we can currently access. It is, therefore, particularly relevant
to study the salient features of this theory in a cosmological
context, hoping to be able to observe some of its implications. The aim of this
letter is precisely to address this  issue by considering
the low energy bosonic string action in a cosmological setting where
the metric, dilaton and Yang-Mills fields are
homogeneous and isotropic.

 Four-dimensional string vacua emerging, for instance,
from heterotic string theories, correspond to N=1 non-minimal
supergravity and super Yang-Mills models [1]. The four-dimensional
 low-energy bosonic action arising from string
theory is, at lowest order in $\alpha^\prime$, the string expansion
parameter,  given by

$$ S_B=\int d^4x \sqrt{-g} \left\{ -{R\over 2 k^2} + 2 (\partial
\phi)^2 - e^{-2 k \phi} Tr\left( F_{\mu\nu} F^{\mu\nu}\right) + 4
V(\phi) \right\} ,\eqno(1)$$

\ni
where $k^2=8\pi M_P^{-2}$, $M_P$ being the Planck mass and we allow
for a dilaton potential,
$V(\phi)$. The field strength $F_{\mu\nu}^a$ corresponds to the one of
a Yang-Mills theory with a gauge group G, which is a subgroup of $E_8\times
E_8$ or
Spin(32)/$Z_2$. We have set the antisymmetric tensor field
$H_{\mu\nu\lambda}$ to zero and dropped the $F_{\mu\nu}^a \tilde
F^{\mu\nu a}$ term --- $\tilde F$ being the dual of F\footnote{$^{(1)}$}{Notice
that, in these models, the dilaton appears mixed with the breathing mode
associated with the radius of the 6-dimensional manifold, in a
combination usually referred to as S and T fields; more general
compactifications may involve a set of $T_i$ fields, the moduli. The
$F\tilde F$ term couples to the imaginary parts of these fields.}.

The theory (1) has been studied in various contexts and, in
particular, black hole and sphaleronic type  solutions [2] have
been found. As we are interested in a cosmological setting,  we shall focus on
homogeneous and isotropic
field configurations on a spatially flat spacetime. The most general
metric is then given by

$$ds^2=-N^2(t) dt^2 + a^2(t) d\Omega^2_3 ,\eqno(2)$$

\ni
where N(t) and a(t) are respectively the lapse function and the scale
factor.

As for the gauge field, we consider for simplicity the gauge group
G=SO(3);  most of our conclusions, however, will be independent of
this choice. An homogeneous and isotropic ansatz, up to a gauge
transformation, for the gauge potential is the following [3]

$$ A_\mu (t) dx^\mu = \sum_{a,b,c=1}^{3} {\chi_0(t)\over 4} T_{a b}
\epsilon_{a c b } dx^c,\eqno(3)$$

\ni
$\chi_0(t) $ being an arbitrary function of time and $T_{ab}$ the generators
of SO(3).

The equations of motion of the Einstein-Yang-Mills-dilaton (EYMD) system for
homogeneous and isotropic field configurations
 can be obtained extremizing (1) with respect to the gauge
potential $A_\mu$, metric $g_{\mu\nu}$ and generic dilaton $\phi$ and
then substitute ans\"atze (2) and (3). In
this work, however, we shall instead follow the equivalent but simpler
procedure
of first dimensionally reduce or consistently truncate the EYMD system allowing
only for homogeneous and isotropic field configurations and only then extremize
with respect to
this subclass of fields to obtain their equations of motion.
This procedure for the Einstein-Yang-Mills
system with SO(N) and SU(M) --- $
N\geq 3, M\geq 2 $ --- gauge groups has been originally developed to study
wormhole type
solutions in a $R\times S^3$ topology [4].

Introducing the ans\"atze (2) and (3) into the action (1) leads, after
integrating over $R^3$ and dividing by the infinite volume of its
orbits, to the following effective action

$$ S_{eff}=-\int_{t_1}^{t_2} dt \left\{ -{3 {\dot a}^2 a \over k^2 N}
+ {3 a\over N} e^{- 2 k \phi} \left[ {{\dot\chi_0}^2\over 2} - {N^2
\over a^2} {\chi_0^4\over 8}\right] + { 2 a^3\over N} {\dot \phi}^2 - 4
a^3 N V(\phi)\right\},\eqno(4)$$

\ni
where the dots denote time derivatives.
The equations of motion, in the gauge N=1, are obtained extremizing
$S_{eff}$ with respect to $a,\ \phi$ and $\chi_0$

$$\eqalignno{
             2 {\ddot a \over a} + H^2 + {k^2\over 3} e^{-2k\phi}
\rho_{\chi_0} + 2 k^2  [ {\dot \phi}^2 - 2 V(\phi)] & = 0,  & (5) \cr
 \ddot \phi + 3 H {\dot \phi} + {1\over 2} k e^{- 2 k \phi} \zeta_{\chi_o} +
{\partial
V(\phi) \over \partial \phi} & = 0, & (6)\cr
  {\ddot \chi_0} + (H-2 k \dot \phi) {\dot \chi_0} + {\chi_0^3 \over 2 a^2} &
=0, & (7) \cr}$$

\ni
where $H=\dot a / a$ and

$$\eqalignno{ \rho_{\chi_o} & = 3  \left[
  {{\dot \chi_0}^2\over 2 a^2} + {\chi_0^4 \over 8 a^4} \right], & (8)\cr
   \zeta_{\chi_0} & =  3      \left[
{{\dot \chi}_0^2\over 2 a^2} - {\chi_0^4 \over 8 a^4} \right]. &(9)\cr
                     } $$

Furthermore,  extremizing
action (4) with respect to N, one obtains  a constraint, Friedmann equation

$$\eqalignno{ H^2 &= {k^2\over 3} \left[ 4 \rho_\phi + e^{-2k\phi}
\rho_{\chi_o}\right]; &(10)\cr
         \rho_\phi & = {1\over 2} {\dot \phi}^2 + V(\phi). &(11)\cr
                }$$

Before discussing a possible inflationary scenario emerging from our
setting, let us comment on a distinct feature of the system of equations
(6) -- (11). Our construction does allow
us to describe radiation through the field $\chi_o$ rather than
treating it as a
macroscopic fluid, a fact which has an immediate bearing on the issue
of energy exchange between the dilaton and the Yang-Mills field. In
fact, it is
then easy to see, working out eqs. (6) -- (11), that

$$\eqalignno{{\dot \rho}_{\phi} & = - 3 H  {\dot
\phi}^2-  {1\over 2} k e^{- 2 k \phi} \zeta_{\chi_o}
    {\dot \phi}, & (12)\cr
     {\dot \rho}_{\chi_0} & = -4 H \rho_{\chi_0}  +
   6 k {\dot \chi_0^2\over a^2} \dot \phi . &(13)\cr} $$

\ni

 The new and somewhat surprising feature  of eqs. (12), (13) is the
appearance of terms proportional to $\dot \phi$. Clearly, these terms do not
play any role in the
reheating process (just as in the case of N=1 supergravity models, in
which chiral superfields are not canonically normalized and where
energy exchange proceeds through an extra ${\dot
\phi}^3$ term [5]). Actually, it is the terms due to $\phi$ decay and
conversion into radiation that, in inflationary models, are
responsible for the reheating process as
the scalar field, the inflaton, quickly oscillates around the minimum of
its potential [6] (in the model studied in refs. [7] and [8], chaotic
inflation is driven by the dilaton itself). In a realistic particle
physics model, $\phi$ decay would occur either through fermionic or bosonic
two-body decays or through $\phi\rightarrow \gamma \gamma $ via
triangular diagrams [8, 9]; this amounts to  introducing
a term proportional to $\Gamma_\phi {\dot \phi}^2$ in eqs. (12), (13), where
$\Gamma_\phi$ is the
$\phi$ field decay width,  given essentially by $\Gamma_\phi \simeq
{m^3\over M_P^2}$ [8, 9] in
the case where the $\gamma \gamma $ mode is dominant.
Terms proportional to $\dot \phi$ may, however, be of relevance
 to the so-called Polonyi problem in 4-dimensional
string models [10] or in models where supersymmetry is broken
dynamically [11] (see below). An energy exchange
of this form was first encountered when coupling minimally
homogeneous and isotropic gauge fields to a multiplet of scalar fields
[3]; actually,  its origin is ultimately related with the coupling of the
dilaton to the kinetic energy terms of other fields.

Let us now turn to the discussion of inflation. As shown in ref. [7],
where  radiation is treated as a fluid, one obtains chaotic inflationary
solutions driven by the dilaton for   $ V(\phi)= {1\over 2} m^2 (\phi -
\phi_0)^2$, with $ 10^{-8}M_P < m < 10^{-6}M_P$ and $\phi_0=M_P$, a
result which
remains valid if we add a quartic term to the potential,
${\lambda\over 4}(\phi - \phi_0)^4$, where bounds on $\lambda$
can be obtained from the condition that energy density perturbations
are not exceedingly large. Although
the dilaton potential, which has its origin is non-perturbative
effects such as gaugino condensation and a possible v.e.v. for the
antisymmetric tensor field [12], has a more complicated structure, it
is reassuring to see that it is possible to obtain inflationary  solutions in
simple cases \footnote{$^{(2)}$}{A quite different inflationary
scenario emerges when accounting for the duality-type string
symmetries which lead to a singularity-free pre-big-bang phase of
accelerated contraction followed, in principle, by a standard
cosmological scenario, free of initial condition problems [13]. }. More
exhaustive analysis lead, however, to less optimistic
conclusions [14] (see also [15]). With our field treatment of radiation
let us repeat the analysis of ref. [7]. We consider the situation
where $H \gg 2 k \dot \phi$, which
allows us to solve eq. (8) in the conformal time $d\eta  = a^{-1}(t) dt$,
the solution being given implicitly in terms of an elliptic function
of the first kind [16]. Furthermore, we find that $\rho_{\chi_0}={C\over
a^4(t)}$,
where $C$ is an integration constant. Substituting these results into
eqs. (6) and (7), we obtain, after introducing the dimensionless
variables $x\equiv {m\over \mu } (\phi - \phi_0),\ y\equiv {1\over \mu} \dot
\phi,\ z\equiv {1\over m} H$ and $\eta\equiv m t$, where $\mu^2={3 m^2 / 2
k^2}$, the
following non-autonomous three-dimensional dynamical system

$$\eqalignno{ x_\eta & = y, &(14-a) \cr
              y_\eta & = - x - 3yz - C_1 {\zeta_{\chi_0}}(t)
                           e^{-\sqrt{6} x}, &(14-b)\cr
              z_\eta &= 2 x^2-y^2-2 z^2, &(14-c)\cr} $$

\ni
where the index $\eta$ denotes derivative with respect to $\eta$ and
we have set $C_1={\sqrt{3\over 2}}  {e^{-2k\phi_0}\over 2 \mu^2}$.  The phase
space of the system is the
region, in $R^3$, characterized  by the constraint equation (11),
which, in the new variables, reads

$$z^2 -x^2 -y^2= \sqrt{2\over 3 }  C_1 {C\over a^4} e^{-\sqrt{6} x}\
.\eqno(15)$$

As argued in refs. [3] and [7], during inflation, terms proportional to
$a^{-4}(t)$ in \break (14-b), where $\zeta_{\chi_0}\sim a^{-4}$, and (15)
become much smaller than the remaining
ones and are therefore negligible (actually, their effect  was
shown to enhance inflation in a few percent [3]). Dropping
these terms, the resulting dynamical system is the very one
encountered in ref. [3], where it is found that there are the
following critical points:

\item{i)} In the finite region of variation of
$x,\ y,\ z$: the origin, $(0,0,0)$, an
asymptotically stable focus.

\item{ii)} In the infinite region, $x^2+ y^2+z^2=\infty$, using spherical
coordinates $(r,\theta,\phi):\ P(\infty, \theta=0),\ S_1(\infty,
\theta=\pi/4, \Psi=0),\ K_1=(\infty, \theta=\pi/4, \Psi=\pi/2),\
S_2=(\infty, {\theta=\pi/ 4}, \Psi=\pi)$ and $ K_2=(\infty,
\theta=\pi/4, \Psi=\pi/2)$, which correspond to saddle (P), sink
($S_1$ and $S_2$)
and source points ($K_1$ and $K_2$), respectively.

 Inflationary solutions do exist and inflation  with more than 65
e-foldings requires that the initial value of the $\phi$ field is such
that $\phi_{i}\gaq 4 M_P$ [3,\ 7] and, actually,  these correspond
  to most of the trajectories, with a probability $1-(m/M_P)^2$.
Inflation is therefore a fairly general feature of models with
$V(\phi)= {1\over 2}m^2 (\phi-\phi_0)^2$ where the initial value of
$\phi$ satisfies the abovementioned condition.

Let us now consider the so-called entropy crisis and Polonyi problems
associated with the
EYMD system. The former  difficulty concerns the dilution of the baryon
asymmetry generated prior to $\phi$ conversion into radiation. In
models where the dilaton mass is very small, such that its lifetime is
greater than the age of the Universe ($\Gamma_\phi^{-1}\geq t_U \approx
10^{60}\ M_P^{-1}$), one may  encounter the Polonyi problem, i.e.
$\rho_\phi$ dominates the energy density of the Universe at present [17].
These problems exist, in particular, in various N=1
supergravity models with one [18] or more [19] chiral superfields and
even non-minimal models [5] as well as in string models [8, 10] and
in dynamical supersymmetry breaking scenarios [11].

 The entropy crisis
problem can be solved either by regenerating the baryon asymmetry
after $\phi$ decay [20, 21] or, as discussed in [8], by considering
models in which the Affleck-Dine mechanism can be implemented to
generate an ${\cal O}(1)$ baryon asymmetry and then allow for its dilution
via $\phi$ decay.

 In what concerns avoiding the
Polonyi problem, a necessary requirement is that, at the time when $\phi$
becomes non-relativistic, i.e. $H(t_{NR})=m$, the ratio of its energy density
to the one
of radiation satisfies [5]

$$ \epsilon={\rho_\phi(t_{NR})\over
\rho_{\chi_0}(t_{NR})}\laq 10^{-8}.\eqno(16)$$

\ni
Notice that, since the condition  $\Gamma_\phi^{-1}\geq t_U$ implies
$m\leq 10^{-20} M_P$, which falls outside the mass interval for which
inflation takes place (see above), we have to assume that, in models
where this problem occurs,  some  field
other than the dilaton will drive inflation and be responsible for
reheating.
In the absence of  the extra
exchange terms proportional to $\dot \phi$ in equations (12) and (13),
initially and until $\phi$ becomes non-relativistic, $\rho_\phi\simeq
\rho_{\chi_0} \simeq {1\over 2} m^2 \phi_\ast^2$, implying that
$\epsilon = {\cal O} (1)$ (see e.g. ref. [5]). Hence,  any mechanism for
draining $\phi$ energy into radiation has to be quite effective in
order to be able to help to avoid the  Polonyi problem.
Let us then estimate the efficiency of the terms proportional to $\dot \phi$
in eqs. (12) and (13).  We get for $\epsilon$

$$\epsilon \simeq {1- 2\Delta /m^2 \phi_\ast^2 \over 1+ 2\Delta^\prime / m^2
\phi_\ast^2},\eqno(17)$$

\ni
where $\phi_\ast\simeq \phi(t_{RN})\approx M_P$ and  $\Delta$, $\Delta^\prime$
are the integrated contributions of the last two terms
of eqs. (12) and (13), respectively,  over the time interval ($t_i,\ t_{NR}$).
One expects
that $\Delta\simeq \Delta^\prime$. Demanding $\epsilon$ to
satisfy condition (16) implies that the ratio $\alpha\equiv {2 \Delta
\over m^2 \phi_\ast^2}$ has to be fairly close to 1. In
order to analyze the implications of this condition, we first notice
that effective energy exchange occurs, as can be seen from eq. (8),
during the period where $H\simeq 2 k \dot \phi$. Assuming that this
relation holds after inflation and, furthermore, that
$\zeta_{\chi_0}\sim a^{-4}$                       and
$a(t)=  a_R \left({t\over t_R}\right)^{1/2}$, we obtain

$$\Delta \simeq {t_R^2\over a^4_R}(t_i^{-2} - t_{NR}^{-2}) ,\eqno(18)$$

\ni
where the index R refers to the time when the inflaton decays.
Hence, in order to get $\alpha={\cal O}(1)$, we must have, if
$t_{NR}\gg t_i$

$$t_i\simeq {1\over m M_P}{t_R\over a_R^2}.\eqno(19)$$

\ni
 For typical values of the relevant parameters, e.g. $t_i\simeq 10^{10}
M_P^{-1}$, $t_R\gaq 10^{30} M_P^{-1}$ and $ a_R\gaq 10^{30}
M_P^{-1}$, we see that the dilaton mass is required to be exceedingly
small, \break $m\leq 10^{-40} M_P$, if the dilaton energy is to be
effectively drained. As discussed above, solving the
Polonyi problem requires $\alpha$ to be very close to 1; it is clear,
from our estimate, that this can be achieved provided the energy
exchange is effectively maintained over a sufficiently long period of time..
 Actually,   energy exchange via terms
proportional to $\dot \phi$  occurs also when coupling the dilaton to
bosons and fermions through $e^{- 2
k \phi} {\cal L}_{matter}$ (see e.g. [7]), which will then contribute to
further draining of $\phi$ energy. Other contributions to this process
would occur if we had chosen a larger gauge
group as, besides $\chi_0(t)$, another multiplet of fields would
appear in the effective action [3, 4] leading to extra   energy
exchange terms.

Notice that, when discussing a very light or massless dilaton, one has
to deal with the implications of the fact that coupling constants and
masses are dilaton dependent and the ensued problems, such as
cosmological variation of the fine structure constant as well as other
coupling constants and violations of the weak equivalence principle.
The study of the cosmological evolution of the Einstein-Matter-Dilaton
system as carried out in ref. [22] indicates that the inclusion of
non-perturbative string loop effects is crucial to render
consistency with the experimental data. The impact of the string loop
effects in the EYMD system is to change the Yang-Mills-dilaton coupling to
$B(\phi)
F_{\mu\nu}^a F^{\mu\nu a}$, where $B(\phi)=e^{-2k\phi} + c_0 + c_1
e^{2k\phi} + c_2 e^{4k\phi} + ...$ and $c_0,\ c_1,\ c_2,...$ are
constants. Hence, in what concerns the Polonyi problem we have been
discussing, the extra  terms imply that the energy transfer
may be actually reversed from the Yang-Mills field to the dilaton,
thereby invalidating our previous conclusions regarding a possible
solution to this problem in the case of an extremely light dilaton.

In summary, we have performed a study of the EYMD system emerging from
string theories. Considering the gauge group $G=SO(3)$, which allows
for the YM field to be completely characterized by a single scalar field, we
have seen
that, as in the description where radiation is regarded as a
fluid (see e.g. ref. [7]), inflationary solutions can be obtained for
\break $V(\phi)={1\over 2} m^2 (\phi - \phi_0)^2$.
Furthermore, a fundamental field theory treatment turns out to be
particularly interesting in understanding the energy balance between
dilaton and radiation as a new exchange term is found which can  help
to avoid
the Polonyi problem in models where the dilaton is light, a
conclusion that can however be invalidated if string loop effects are included.

\vskip 0.4cm

{\it Acknowledgements:}
\vskip .2cm
 One of us (O. B.) would like to thank Prof. G.
Veneziano for important discussions.

\vfill
\eject
\centerline{\bf References}
\vskip 0.3cm

\ni
\item{[1]} E. Witten, Phys. Lett. B155 (1985) 151; Nucl. Phys. B268
(1986) 79.

\ni
\item{[2]} D. Garfinkle, G. T. Horowitz and A. Strominger, Phys. Rev.
D43 (1991) 3140;

E.E. Donets and D. V. Gal'tsov, Phys. Lett. B302 (1993) 411;

 G. Lavrelashvili and D. Maison, Nucl. Phys. B410 (1993) 407.

\ni
\item{[3]} P.V. Moniz, J.M. Mour\~ao and P.M. S\'a, Class. Quantum
Gravity 10 (1993) 517.

\ni
\item{[4]} O. Bertolami, J. M. Mour\~ao, R.F. Picken  and I. P. Volobujev, Int.
J.
Mod. Phys. A6 (1991) 4149.

\ni
\item{[5]} O. Bertolami, Phys. Lett. B209 (1988) 277.

\ni
\item{[6]} A. Albrecht, P.J. Steinhardt, M.S. Turner and F. Wilczek,
Phys. Rev. Lett. 48 (1982) 1427;

M. Morikawa and M. Sasaki, Prog. Theor. Phys. 72 (1984) 782;

A. Ringwald, Ann. Phys. (NY) 177 (1987) 129.

\ni
\item{[7]} M.C. Bento, O. Bertolami and P.M. S\'a, Phys. Lett. B262
(1991) 11.

\ni
\item{[8]} M.C. Bento, O. Bertolami and P.M. S\'a, Mod. Phys. Lett.
A7 (1992) 911.

\ni
\item{[9]} M. Yoshimura, Phys. Rev. Lett. 66 (1991) 1559.

\ni
\item{[10]} B. de Carlos, J.A. Casas, F. Quevedo and E. Roulet, Phys.
Lett. 318 (1993) 447.

\ni
\item{[11]} T. Banks, D. B. Kaplan and A. E. Nelson, Preprint UCSD/PTH
93-26, RU-37(1993).

\ni
\item{[12]} M. Dine, R. Rohm, N. Seiberg and E. Witten, Phys. Lett.
B156 (1985) 55.

\ni
\item{[13]} M. Gasperini and G. Veneziano, Ast. Phys. 1 (1993) 317.

\ni
\item{[14]} P. Binetruy and M.K. Gaillard, Phys. Rev. D34 (1986) 3069.

\ni
\item{[15]} R. Brustein and P.J. Steinhardt, Phys. Lett. B302 (1993)
196.

\ni
\item{[16]} M.C. Bento, O. Bertolami, P.V. Moniz, J.M. Mour\~ao and
P.M. S\'a, Class. Quantum Gravity 10 (1993) 285.

\ni
\item{[17]} G.D. Coughlan, W. Fischler, E.W. Kolb, S. Raby and G.G.
Ross, Phys. Lett. B131 (1983) 59;

J. Ellis, K. Enqvist and D.V. Nanopoulos, Phys. Lett. B151 (1985) 357.

\item{[18]} G.D. Coughlan, R. Holman, P. Ramond and G.G. Ross, Phys.
Lett. B140 (1984) 44.

\item{[19]} O. Bertolami and G.G. Ross, Phys. Lett. B171 (1986) 46.

\ni
\item{[20]} V. Kuzmin, V. Rubakov and M. Shaposhnikov, Phys. Lett.
B155 (1985) 36.

\ni
\item{[21]} M. Claudson, L.J. Hall and I. Hinchcliffe, Nucl. Phys.
B241 (1984) 309;

K. Yamamoto, Phys. Lett. B168(1986)341;

O. Bertolami and G.G. Ross, Phys. Lett. B183 (1987) 163;

A. Cline and S. Raby, Phys. Rev. D43 (1991) 1781;

S. Mollerach and E. Roulet, Phys. Lett. B281 (1992) 303.

\item{[22]} T. Damour and A. M. Polyakov, HEPTH-9401069.

\bye